\documentclass[journal=jacsat,manuscript=article]{achemso}

\usepackage[version=3]{mhchem} 

\author{KeYuan Ma}
\affiliation[University of Zurich]
{Department of Chemistry, University of Zurich, CH-8057 Z\"urich, Switzerland}

\author{Karolina Gornicka}
\affiliation [Gdansk University of Technology]
{Faculty of Applied Physics and Mathematics, Gdansk University of Technology, Gdansk 80-233, Poland}

\author{Robin Lef\`evre}
\affiliation[University of Zurich]
{Department of Chemistry, University of Zurich, CH-8057 Z\"urich, Switzerland}

\author{Yikai Yang}
\affiliation[Ecole Polytechnique F\'ed\'erale de Lausanne (EPFL)]
{Institute of Physics, Ecole Polytechnique F\'ed\'erale de Lausanne (EPFL), CH-1015 Lausanne, Switzerland}

\author{Henrik M. R\o nnow}
\affiliation[Ecole Polytechnique F\'ed\'erale de Lausanne (EPFL)]
{Institute of Physics, Ecole Polytechnique F\'ed\'erale de Lausanne (EPFL), CH-1015 Lausanne, Switzerland}

\author{Harald O. Jeschke}
\affiliation[Okayama University]
{Research Institute for Interdisciplinary Science, Okayama University, Okayama 700-8530, Japan}

\author{Tomasz Klimczuk}
\affiliation [Gdansk University of Technology]
{Faculty of Applied Physics and Mathematics, Gdansk University of Technology, Gdansk 80-233, Poland}

\author{Fabian O. von Rohr}
\affiliation[University of Zurich]
{Department of Chemistry, University of Zurich, CH-8057 Z\"urich, Switzerland}

\title[An \textsf{achemso} demo]
  {Superconductivity with High Upper Critical Field in the Cubic Centrosymmetric $\eta$-Carbide \ce{Nb4Rh2C_{1-\delta}}}


\begin{document}

\begin{tocentry}
\includegraphics[]{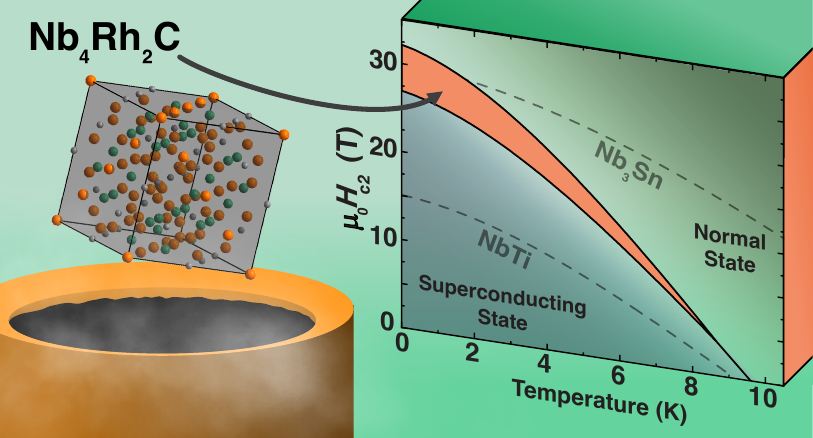}

\end{tocentry}

\begin{abstract}
  The upper critical field is a fundamental measure of the strength of superconductivity in a material. It is also a cornerstone for the realization of superconducting magnet applications. The critical field arises because of the Copper pair breaking at a limiting field, which is due to the Pauli paramagnetism of the electrons. The maximal possible magnetic field strength for this effect is commonly known as the Pauli paramagnetic limit given as $\mu_0 H_{\rm Pauli} \approx 1.86{\rm [T/K]} \cdot T_{\rm c}$ for a weak-coupling BCS superconductor. The violation of this limit is only rarely observed. Exceptions include some low-temperature heavy fermion and some strongly anisotropic superconductors. Here, we report on the superconductivity at 9.75 K in the centrosymmetric, cubic $\eta$-carbide-type compound \ce{Nb4Rh2C_{1-\delta}}, with a normalized specific heat jump of $\Delta C/\gamma T_{\rm c} =$ 1.64. We find that this material has a remarkably high upper critical field of $\mu_0 H_{\rm c2}{\rm (0)}$ =~28.5~T, which is exceeding by far its weak-coupling BCS Pauli paramagnetic limit of $\mu_0 H_{\rm Pauli}$~=~18.1 T. Determination of the origin and consequences of this effect will represent a significant new direction in the study of critical fields in superconductors.
\end{abstract}

\section{Introduction}
The discovery of new superconducting materials with improved superconducting properties -- for a wide-range of superconducting applications -- is a long-standing challenge in solid-state chemistry\cite{canfield2011}. One of the most important properties for superconducting applications is the upper critical field $H_{\rm c2}$. Superconducting magnets are of fundamental technological importance e.g. for magnetic resonance imaging (MRI) and nuclear magnetic resonance (NMR) applications\cite{Blatter1994,Moll2010,hahn2019}. When an external magnetic field acts on a superconducting material, there are two effects that may cause a Cooper pair breaking, (i) the orbital-limiting effect and (ii) the Pauli paramagnetic effect, also called Zeeman effect.\cite{Tinkham2004} The orbital-limiting effect causes a breaking of pairs by inducing a momentum in the single-particle spectrum that leads to a supercurrent, which exceeds the superconducting gap. The Pauli paramagnetic effect corresponds to the Zeeman energy of the electron in the single-particle spectrum that, if large enough, exceeds the superconducting condensation energy. The orbital-limiting effect is commonly the dominant pair breaking effect close to the critical temperature $T_{\rm c}$ of a superconductor, while at low temperatures, far away from the critical temperature $T_{\rm c}$, the Pauli paramagnetic effect is dominant\cite{Tinkham2004}. 
The maximal upper critical field $H_{\rm c2}$ in the weak-coupling Bardeen-Schrieffer-Cooper (BCS) theory for superconductivity is given by the paramagnetic pair breaking effect, which is commonly known as the Pauli paramagnetic limit $H_{\rm Pauli}$\cite{Tinkham2004}. This value is given by 

\begin{equation}
    \mu_0 H_{\rm Pauli} = \frac{\Delta_0}{\sqrt{2} \mu_{\rm B}} \approx 1.86{\rm [T/K]} \cdot T_{\rm c}
\end{equation}
with $\mu_{\rm B}$ being the Bohr magneton and $\Delta_0$ being the superconducting gap. 

The Pauli paramagentic limit can be higher for strong-coupled superconductors \cite{clogston1962upper,orlando1979critical}. The violation of the Pauli paramagnetic limit is of great interest for exotic quantum states \cite{Bristow2020,Hunte2008}. When the orbital pair breaking effect is suppressed, novel superconducting states are expected to appear such as for example the so-called Fulde-Ferrell-Larkin-Ovchinnikov (FFLO) state, where Cooper pairs with finite momentum are formed across a split Fermi surface \cite{fulde1964superconductivity,buzdin1997generalized,Gurevich2010,Song2019}. The weak-coupling Pauli paramagnetic limit commonly holds for almost all type-II superconductors. However, there are still some superconductors that violate it. Cases where Pauli paramagnetic limit violations are being discussed include (i) highly anisotropic superconductors\cite{tanaka2020superconducting,lu2014superconductivity,lei2012multiband,falson2020type}, (ii) Chevrel phases\cite{Morley2001,petrovic2011multiband}, (iii) heavy fermion superconductors \cite{bauer2004heavy,manago2017absence}, and (iii) the spin-triplet superconductor \ce{UTe2}\cite{aoki2019unconventional}. Generally, the expected causes for this violation is observed for special combinations of electronic and structural properties. It can be caused by large coupling of the electron spins and the spin-orbit coupling, or by an unconventional superconducting pairing process.

The upper critical fields $\mu_0 H_{\rm c2}(0)$ of superconductors are technologically highly relevant, as they are limiting the achievable magnetic fields in a superconducting magnet. The majority of superconducting magnets remain today made of intermetallic compounds, especially of NbTi and \ce{Nb3Sn} wires.\cite{geballe1993superconductivity,sharma2015superconductivity} NbTi wires are the only known ductile superconductors with an $\mu_0 H_{\rm c2}(0) > 10$ T. This alloy with the ideal composition of Ti-Nb (32 at. \%) has a critical temperature of $T_{\rm c} \approx$ 8.5 K and an upper critical field of $\mu_0 H_{\rm c2}(0) \approx$ 15.5 T.\cite{collings2012sourcebook} Superconducting magnets for higher-field applications are most commonly made containing \ce{Nb3Sn}. \ce{Nb3Sn} has a critical temperature of $T_{\rm c} \approx $ 18 K and an upper critical field of $\mu_0 H_{\rm c2}(0) \approx$ 30 T.\cite{sharma2015superconductivity} The brittle nature of \ce{Nb3Sn} made it necessary to develop stabilized multifilamentary composites in order to produce wires and superconducting magnets out of this material. For both superconductors NbTi, as well as, \ce{Nb3Sn} the upper critical fields $H_{\rm c2}(0)$ is below the Pauli paramagnetic limit $H_{\rm Pauli}$.

Here, we report on the bulk superconductivity and the remarkably high-upper critical fields in the centrosymmetric, cubic $\eta$-carbide \ce{Nb4Rh2C_{1-\delta}}. This material displays a critical temperature of $T_{\rm c} =$ 9.7 K and a normalized specific heat jump $\Delta C/\gamma T_{\rm c}$ of 1.64, which is close to the weak-coupling BCS value of 1.43. The upper-critical field in this material is with $\mu_0 H_{\rm c2}(0) =$ 28.5 T exceeding by far the weak-coupling BCS Pauli paramagnetic limit of $\mu_0 H_{\rm Pauli} =$ 18.1 T. This violation occurs in this bulk superconductor in the absence of any obviously exotic electronic properties.

\section{Experimental}

{\it Synthesis.-} The starting materials for the synthesis of \ce{Nb4Rh2C_{1-\delta}} samples were niobium powder (99.99 \%, -325 mesh, Alfa Aesar), rhodium powder (99.95 \%, Sigma-Aldrich), and carbon (99.999 \%, Sterm Chemicals). Samples of a mass of 150 mg of the starting materials Nb, Rh, and C were weighed in nearly stoichiometric ratios. Highest purity samples were obtained for a Nb:Rh:C ratio of 3.9:2.1:0.7. The reactants were thoroughly mixed and pressed into a pellet ($d$ = 0.8 cm, with 3-ton pressure). The pellet was melted in an arc furnace in a purified argon atmosphere on a water-cooled copper plate. The sample was molten 10 times in order to ensure optimal homogeneity of the elements. After arc melting, a negligible small mass loss of 2-3 \% was observed. The solidified melt was ground to a fine powder and pressed into a pellet ($d$ = 0.8 cm, with 3-ton pressure). The pellet was sealed in a quartz ampule under 1/3 atm argon and annealed in a furnace for 7 days at 1100 $^\circ$C. After the reaction the quartz tube  was quenched in water. We observed that any variation from the compositional ratio or from the annealing temperature leads to the formation of impurity phases (see Supporting Information). PXRD check on the solidified melt after arc-melting did not show diffraction peaks from the final product, suggesting the \ce{Nb4Rh2C_{1-\delta}} was formed during the annealing process.      

{\it Analytical methods.-} The crystal structure and phase purity of the \ce{Nb4Rh2C_{1-\delta}}  sample were checked using powder x-ray diffraction (PXRD) measurements on a STOE STADIP diffractometer with Mo K$_{\alpha1}$ radiation ($\lambda$ = 0.70930 Å). The PXRD patterns were collected in the 2$\Theta$ range of 5-50$^{\circ}$ with a scan rate of 0.25$^{\circ}$/min. Rietveld refinements were performed using the FULLPROF program package. The morphology and composition of the polycrystalline samples were examined under a scanning electron microscope (SEM) equipped with an energy-dispersive X-ray (EDX) spectrometer. The arrangement of atoms and the corresponding selected area electron diffraction pattern (SAED) were observed by a high-resolution transmission electron microscope (HRTEM). Data processing was done using the software ReciPro.    
The temperature- and field-dependent magnetization measurements were performed using a Quantum Design magnetic properties measurement system (MPMS3) with a 7 T magnet equipped with a vibrating sample magnetometry (VSM) option. The measured pellets were placed in parallel to the external magnetic field to minimize demagnetization effects. Specific heat capacity and resistivity measurements were performed with a Quantum Design physical property measurement system (PPMS) Evercool with a 9 T magnet. The standard four-probe technique was employed to measure the electrical resistivity with an excitation current of $I$ = 1.5 mA. In the resistivity measurement, gold wires were connected to the sample with silver paint. Specific heat measurements were performed with the Quantum Design heat-capacity option, using a relaxation technique. The high magnetic field (10-17 T) electronic transport measurements were performed on a superconducting magnet made by Oxford instrument (power supply: IPS-120) with a  dilution fridge (Kelvinox-400).  

{\it Computational methods.-} Electronic structure calculations were performed using the all electron full potential local orbital (FPLO) basis set~\cite{koepernik1999full}. We use both scalar relativistic and fully relativistic calculations in order to check for effects of spin orbit coupling. We use $12\times 12\times 12$ $k$ meshes and fully converge both charge density and energy. We employ the generalized gradient approximation (GGA) to the exchange correlation functional~\cite{perdew1996generalized} and use a GGA+U correction for Rh and Nb $4d$ orbitals~\cite{liechtenstein1995} in order to check for magnetic tendencies. We approximate the \ce{Nb4Rh2C_{1-\delta}} structure by setting $\delta$ to zero. As there are no carbon states at the Fermi level, we expect that this does not affect the electronic structure near the Fermi level.

\section{Results and discussion}

\begin{figure}
\centering
\includegraphics[width=0.7\textwidth]{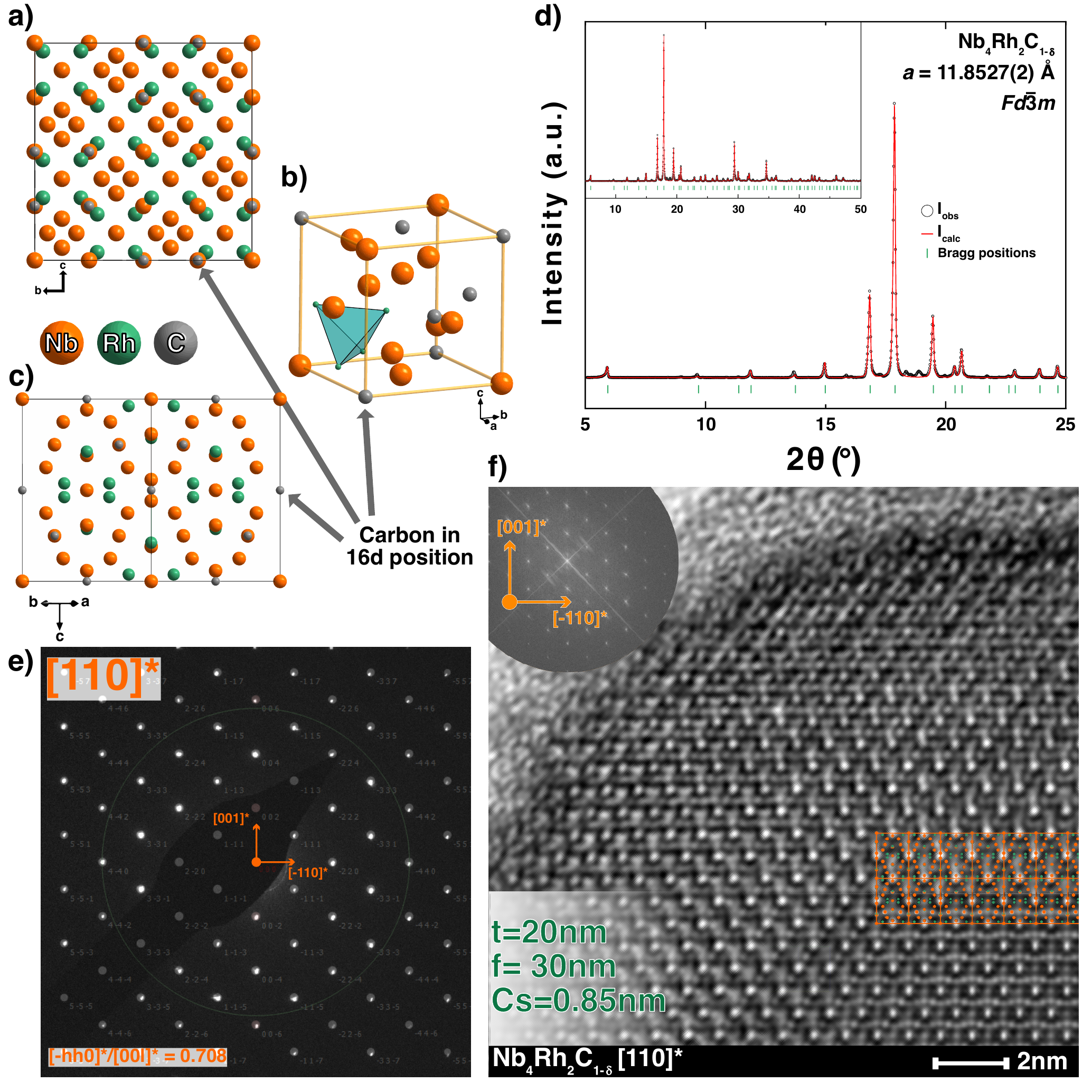}
\caption{Crystal structure of \ce{Nb4Rh2C_{1-\delta}}. (a,b,c) Different crystal directions. (d) Rietveld refinement and PXRD data. (e) Measured [110]* electron diffraction pattern of \ce{Nb4Rh2C_{1-\delta}} superimposed to the theoretical one. (f) The lower figure is a section of a larger experimental [110] oriented HRTEM-image. Upper-left corner: The FFT of the complete HRTEM-image. The lower-right corner is a HRTEM simulation.}
\label{fig:figure1}
\end{figure}

\ce{Nb4Rh2C_{1-\delta}} crystallizes in the $\eta$-carbide structure in the cubic space group $Fd\bar{3}m$ (no. 227)\cite{Westgren1926}. $\eta$-Carbides are sub-carbides, nearly intermetallic materials that form for the compositions $N_4M_2X_{1-{\delta}}$ and $N_3M_3X_{1-{\delta}}$ with \textit{N} and \textit{M} being transition metals, and \textit{X} being carbon, nitrogen, or oxygen \cite{Westgren1926,souissi2018effect,Ma2019,Waki2010,Waki2011}. The crystal structure of ideal \ce{Nb4Rh2C} contains 112 atoms in the unit cell, which are fully described by 4 crystallographic sites. The asymmetric unit is echoed eight times over the whole unit cell following the symmetry elements of the space group. Figure \ref{fig:figure1}(b) represents the asymmetric unit of the full structure shown in figure \ref{fig:figure1}(a)\&(c) along different directions.

The powder x-ray diffraction (PXRD) pattern and the corresponding Rietveld refinement of \ce{Nb4Rh2C_{1-\delta}} are shown in figure \ref{fig:figure1}(d). The cell parameter of \ce{Nb4Rh2C_{1-\delta}} was determined to be $a_0 =$ 11.8527(2) \AA. In figure \ref{fig:figure1}(e)\&(f) electron diffraction and HRTEM images of \ce{Nb4Rh2C_{1-\delta}} are shown. We observe the highly ordered Nb and Rh columns of atoms along the [110] direction. The fast Fourier transform (FFT) process of the HRTEM image shows the high order present in the crystal. The observed distances are in very good agreement with the superimposed crystal structure. The corresponding selected area electron diffraction displays a set of well-arranged diffraction spots, exemplifying the high crystallinity of the sample. The composition has been determined from energy-dispersive X-ray spectroscopy (EDX) for Nb:Rh to be 2.03(4) on the same sample. 

\begin{figure}
\centering
\includegraphics[width=0.5\linewidth]{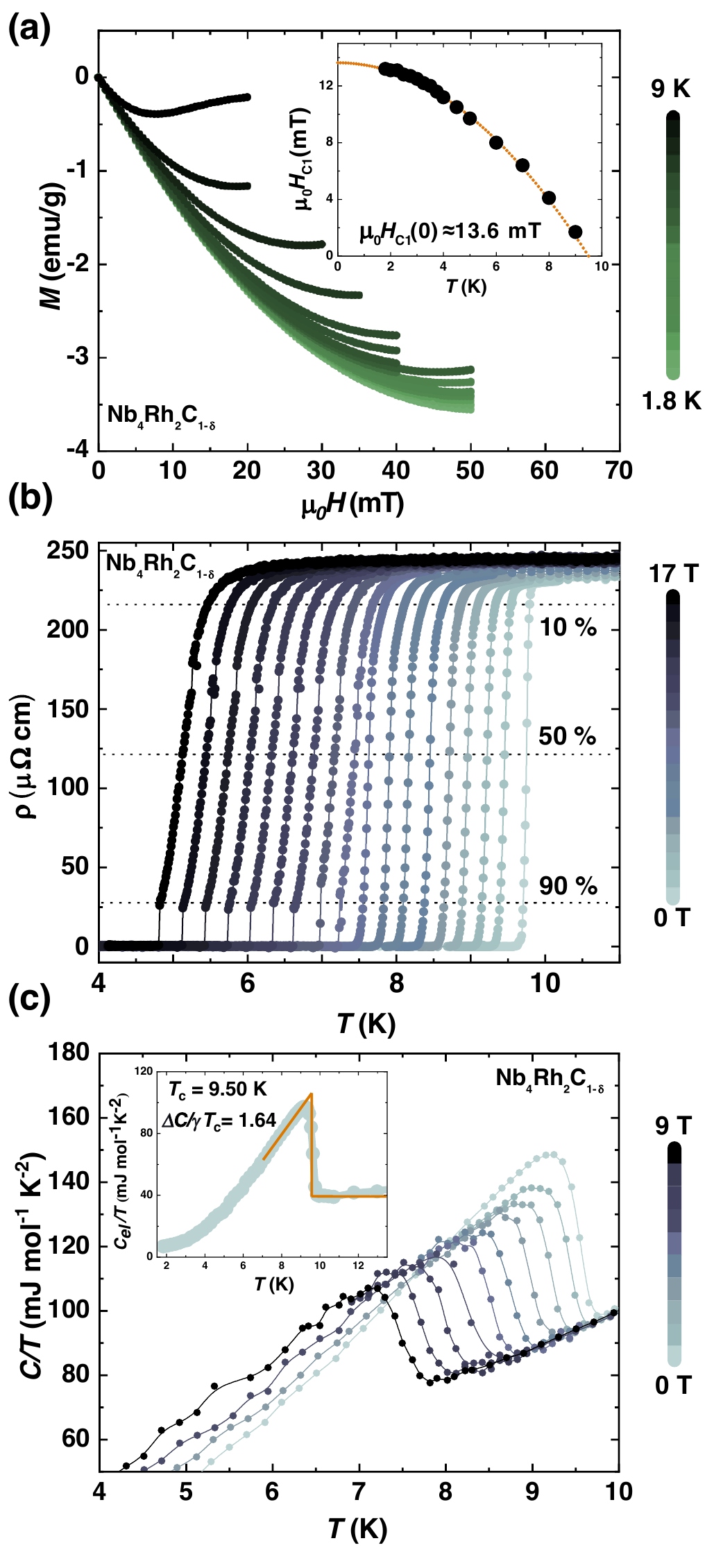}\\
\caption{Superconducting properties of \ce{Nb4Rh2C_{1-\delta}}. (a) ZFC magnetization $M$($H$) in a temperature range between $T$ = 1.8 K and 9 K for external fields between $\mu_0 H =$ 0 T and 50 mT. Inset: Lower-critical field $H_{\rm c1}$($T$), orange line is a fit using the empirical formula: $H_{c1}(T) = H_{c1}(0) [1-(T/T_c)^2]$. (b) Field-dependent resistivity in the vicinity of the superconducting transition in fields between $\mu_0 H =$ 0 T and 17 T. (c) Field dependent specific heat in fields between $\mu_0 H =$ 0 T and 9 T. Inset: Entropy-conserving construction of the zero-field measurement.}
\label{fig:figure2}
\end{figure}

We observe a sharp superconducting transition at $T_{\rm c}$  = 9.70 K in the temperature dependence of the magnetic susceptibility in zero-field cooled (ZFC) and field-cooled (FC) modes, respectively (see Supporting Information). In agreement with an earlier report by Ku \textit{et al.}, who reported a $\rm{T}_c \approx$ 9.3 K in the Nb-Rh-C system associated with an $\eta$-carbide phase\cite{ku1985effect,ku1984new}. We find that this material is a Pauli paramagnet in the normal state, as confirmed by magnetization measurement between $T =$ 10 K to 300 K in fields of $\mu_0 H =$ 1 T, 5 T and 7 T (see Supporting Information). A series of field-dependent measurements of the ZFC magnetization in low fields allows us to determine the lower critical field $H_{\rm c1}$, as shown in figure \ref{fig:figure2}(a). Here, we used the magnetic-field point where the $M$($H$) curve first deviates from linearity as the measure for $H_{\rm c1}$ \cite{naito1990temperature}. With this approximation, the obtained $H_{\rm c1}$ values are fitted using the empirical formula
\begin{equation}
H_{c1}(T) = H_{c1}(0) [1-(T/T_c)^2]. 
\end{equation}
The lower critical field at $T =$ 0 K is determined to be $\mu_0 H_{\rm c1}(0) $= 13.6 mT. 

In figure \ref{fig:figure2}(b), we show the resistivity $\rho$($T$,$H$) in a temperature range between $T$ = 4 K and 11 K. The resistivity shows an abrupt drop to zero at the transition to the superconducting state for all fields. In zero-field, we determine the critical temperature to be $T_{\rm c}$ = 9.75 K. As expected, the critical temperature decreases steadily as the applied magnetic field increases. In the maximally applied field of $\mu_0 H =$~17~T the critical temperature in the resistivity is still $T_{\rm c} =$ 5.14 K. Furthermore, it is noteworthy that the transition width $\Delta T_{\rm c}$ is small and the transition, therefore, extremely sharp even in high-fields. This indicates a large vortex-solid region, and a small vortex-liquid region. In the normal-state the resistivity decreases with decreasing temperature in a nearly linear temperature dependence of $\rho = \rho_0 + \beta T$. Here the residual resistivity ratio (RRR) value is defined as $\rho$(300 K)/$\rho$(9.7 K) $\approx$ 1.16, corresponding to a poor metal (see Supporting Information)\cite{von2014,von2017}. This small RRR value may arise from the polycrystalline nature of the sample with grain boundaries and macroscopic defects.

Figure \ref{fig:figure2}(c) shows the specific heat $C$($T$,$H$) in the vicinity of the superconducting transition. The data is plotted as $C/T$ versus $T$ under fields of 0 to 9 T. The specific heat can be fitted according to the expression: 

\begin{equation}
\frac{C(T)}{T} = \frac{C_{el} + C_{ph}}{T} = \gamma + \beta T^2
\end{equation}

We find the Sommerfeld parameter $\gamma$ to be 40 mJ mol$^{-1}$ K$^{-2}$, and the phononic coefficient $\beta$ to be 0.6 mJ mol$^{-1}$ K$^{-4}$ for \ce{Nb4Rh2C_{1-\delta}} (see Supporting Information). We determined the Debye temperature to be $\Theta_D =$ 283 K, according to the following relationship:  

\begin{equation}
\Theta_D = \left(\frac{12 \pi^4}{5 \beta} n R \right)^{\frac{1}{3}} 
\end{equation}
Here n = 7 is the number of atoms per formula unit, and R = 8.314 J mol$^{-1}$ K$^{-1}$ is the ideal gas constant.

The specific heat jump corresponding to the superconducting transition is observed at $T_{\rm c}$ = 9.51 K in zero-field. The normalized specific heat jump, $\Delta C/\gamma T_{\rm c}$ is found to be 1.64 in zero-field, which is slightly larger than the weak-coupling BCS value of 1.43. This value is indicating enhanced-coupling superconductivity in this material. The specific heat jump is well-pronounced in all measured fields, and shifts to lower temperatures in good agreement with the results from the resistivity measurements. 

The electron-phonon coupling constant $\lambda_{\rm ep}$ can be estimated from the Debye temperature, using the semi-empirical McMillan approximation \cite{mcmillan1968transition}: 
\begin{equation}
\lambda_{\rm ep} = \dfrac{1.04 + \mu^{*} \ {\rm ln}\big(\frac{\Theta_{\rm D}}{1.45 T_{\rm c}}\big)}{(1-0.62 \mu^{*}) {\rm ln}\big(\frac{\Theta_{\rm D}}{1.45 T_{\rm c}}\big)-1.04}.
\end{equation} 
The Coulomb repulsion parameter $\mu^{*}$ typically lays in the 0.1 - 0.15 range. Here, $\mu^{*}$ is set to be 0.13 according to the empirical approximation for similar materials\cite{mcmillan1968transition,von2016}. Based on the obtained values, the $\lambda_{\rm ep} $ value is calculated to be 0.83. 

The measured $\gamma$ value is comparatively large, which might indicate strong electronic correlations in \ce{Nb4Rh2C_{1-\delta}}, it corresponds to a density of states at the Fermi-level of $D(E_{\rm F})$ of 9.32 states eV$^{-1}$ per formula unit (f.u.), when using the following relationship:

\begin{equation}
D(E_{\rm F}) = \dfrac{3 \gamma}{\pi^2 k_{\rm B}^2 (1+\lambda_{\rm ep})}.
\end{equation}

\begin{figure}
\centering
\includegraphics[width=0.8\textwidth]{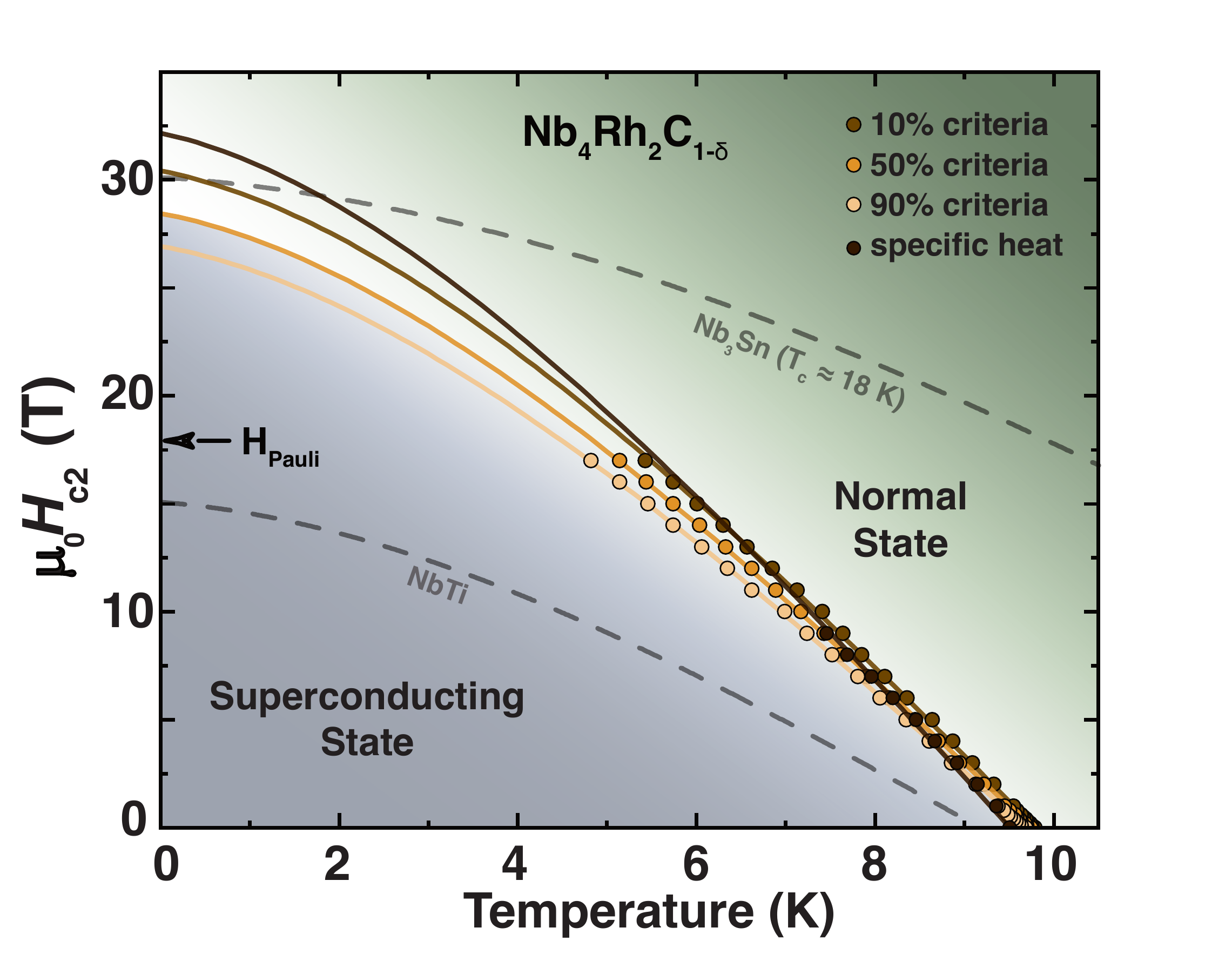}\\
\caption{upper-critical fields $H_{\rm c2}$ of \ce{Nb4Rh2C_{1-\delta}}. Data points from specific heat measurements, and using the 10\%, 50\%, and 90\%-criteria from the resistivity are shown. The data was fitted using equation \ref{WHH}. For comparison, the critical fields of optimal NbTi and \ce{Nb3Sn} alloys are depicted after references \citenum{larbalestier2011high}.}
\label{fig:analysis}
\end{figure}

The field-dependence of $T_{\rm c}$ in the specific heat capacity and the resistivity for the commonly used 10\%-, 50\%-, and 90\%-criteria are depicted in figure \ref{fig:analysis}. The upper-critical field $H_{\rm c2}$(0) can be well fitted using the Werthamer-Helfand-Hohenberg (WHH) formalism in the clean limit according to\cite{baumgartner2013effects}:

\begin{equation}
\label{WHH}
\mu_0 H_{c2} (T)= \frac{\mu_0 H_{c2} (0)}{0.73} h^\ast_{fit} (T/T_{\rm c}).
\end{equation}
with $h^\ast_{fit}$ being
\begin{equation}
 h^\ast_{fit} (t)=(1-t)-C_{1}(1-t)^2-C_{2}(1-t)^4.
\end{equation}
with the parameters $C_{1}$=0.135 and  $C_{2}$=0.134 were used for the fitting. The zero-temperature upper-critical field $\mu_0 H_{\rm c2}$(0) is determined to be 30.4 T, 28.5 T, 26.9 T, and 32.2 T, respectively. 

All values are by far exceeding the weak-coupling BCS Pauli paramagnetic limits of $\mu_0 H_{\rm Pauli} =$ 18.2 T, 18.1 T, 18.0 T, and 17.7 T, respectively. For comparison the critical fields of the two most technologically relevant superconductors NbTi and \ce{Nb3Sn} are depicted in figure \ref{fig:analysis}.\cite{larbalestier2011high} The comparison exemplifies how remarkable the large upper critical field of \ce{Nb4Rh2C_{1-\delta}} is. With a critical temperature similar to the NbTi alloy, this $\eta$-carbide superconductor reaches an upper critical field $\mu_0 H_{\rm c2}$(0) comparable to that of \ce{Nb3Sn}.

The resulting upper critical field from the 50\%-criteria is 28.5 T and it corresponds to a short superconducting coherence length of $\xi_{\rm GL}$ $=$ 34.0 \AA \ according to

\begin{equation}
H_{\rm c2}(0) = \frac{\Phi_0}{2 \pi \ \xi_{\rm GL}^2}.
\label{eq:GL}
\end{equation}

with $\Phi_0 = h/(2e) \approx 2.0678 \times  10^{-15}$ Wb being the quantum flux. The superconducting penetration depth $\lambda_{\rm GL}$ was estimated from the above obtained values of $\xi_{\rm GL}$ and $H_{c1}$ by using the relation:
 
\begin{equation}
\mu_0 H_{c1} = \frac{\Phi_0}{4 \pi \lambda_{\rm GL}^2} ln(\frac{\lambda_{\rm GL}}{\xi_{\rm GL}}).
\end{equation}

We obtained a value of $\lambda_{\rm GL} =$ 2252 \AA \ for \ce{Nb4Rh2C_{1-\delta}}. The value of $\kappa_{\rm GL} = \lambda_{\rm GL}/\xi_{\rm GL} =$ 66.2. Combining the results of $H_{c1}$, $H_{c2}$, and $\kappa_{\rm GL}$, the thermodynamic critical field $\mu_0 H_{c} = $ 304.1 mT can be estimated from the equation:

\begin{equation}
\mathrm{} \quad {H_{c1}}{H_{c2}} = {H_{c}^2} ln(\kappa_{\rm GL}).
\end{equation}

These values demonstrate that \ce{Nb4Rh2C_{1-\delta}} is an extreme type-II superconductor with a short superconducting coherence length and a large superconducting penetration depth $\lambda_{\rm GL}$. All the parameters that we have obtained here for \ce{Nb4Rh2C_{1-\delta}} are summarized in table \ref{tab:super}.

\begin{table} [H]
\caption{Summary of all the determined superconducting parameters of \ce{Nb4Rh2C_{1-\delta}}.}
	\begin{center}
		\begin{tabular}{| c | c |c|}
			\hline
			\ \ Parameters \ \ & \ \    Units\ \ &\ce{Nb4Rh2C_{1-\delta}}\ \ \\
			\hline
			$T_{\rm c,magnetization}$  \ \ & \ \  \ K\ \ & \ \  \ 9.70\ \ \\
			$T_{\rm c,resistivity}$  \ \ & \ \  \ K\ \ & \ \  \ 9.75\ \ \\
			$T_{\rm c,specific heat}$  \ \ & \ \  \ K\ \ & \ \  \ 9.51\ \ \\
			$\rho$(300K)  \ \ & \ \  \ m$\Omega$ cm\ \ & \ \  \ 0.27\ \ \\
			RRR \ \ & \ \  \ -\ \ & \ \  \ 1.16\ \ \\
			$\mu_0 H_{\rm c1}(0)$ \ \ & \ \  \ mT  \ \ & \ \  \ 13.6\ \ \\
			$\mu_0 H_{\rm c2}(0)$  \ \ & \ \  \ T\ \ & \ \  \ 28.5\ \ \\
			$\beta$  \ \ & \ \  \ mJ mol$^{-1}$ K$^{-4}$\ \ & \ \  \ 0.6\ \ \\
			$\gamma$  \ \ & \ \  \ mJ mol$^{-1}$ K$^{-2}$\ \ & \ \  \ 40\ \ \\
			$\Theta_D$  \ \ & \ \  \ K\ \ & \ \  \ 283\ \ \\
			$\xi_{GL}$  \ \ & \ \  \ \AA\ \ & \ \  \ 34\ \ \\
			$\lambda_{GL}$ \ \ & \ \  \ \AA\ \ & \ \  \ 2252\ \ \\
			$\kappa_{\rm GL}$ \ \ & \ \  \ -\ \ & \ \  \ 66.2\ \ \\
			$\mu_0 H_{c}(0)$ \ \ & \ \  \ mT  \ \ & \ \  \ 304.1\ \ \\
			$\Delta C/\gamma T_{\rm c}$ \ \ & \ \  \ -\ \ & \ \  \ 1.64\ \ \\
			$D$($E_{\rm F}$) \ \ & \ \  \ states eV$^{-1}$ per f.u.\ \ & \ \  \ 9.32\ \ \\
			$\mu_0 H_{\rm c2}(0)$/$T_{c}$  \ \ & \ \  \ T/K \ \ & \ \  \ 2.92\ \ \\
			\hline
		\end{tabular}
		\label{tab:super}
	\end{center}
\end{table}

We performed temperature-dependent magnetoresistance and Hall resistance  measurements on \ce{Nb4Rh2C_{1-\delta}} in the normal state (see Supporting Information). At 10 K, we observed a linear magnetic field-dependent magnetoresistance in \ce{Nb4Rh2C_{1-\delta}} between $\mu_0 H =$ 2 and 9 T. The resistance thereby, increases about 0.7 \% at 9 T. At higher temperatures, the magnetoresistance decreases notably. The Hall resistance measurements at different temperatures show negative slope values over magnetic fields. The negative Hall constant values indicate that electrons are the dominant charge carriers in \ce{Nb4Rh2C_{1-\delta}}. The calculated carrier density at 200 K is 2.53 $\times$ 10$^{28}$ m$^{-3}$, which corresponds to 4.2 electrons per formula unit. 

\begin{figure}
\centering
\includegraphics[width=0.6\textwidth]{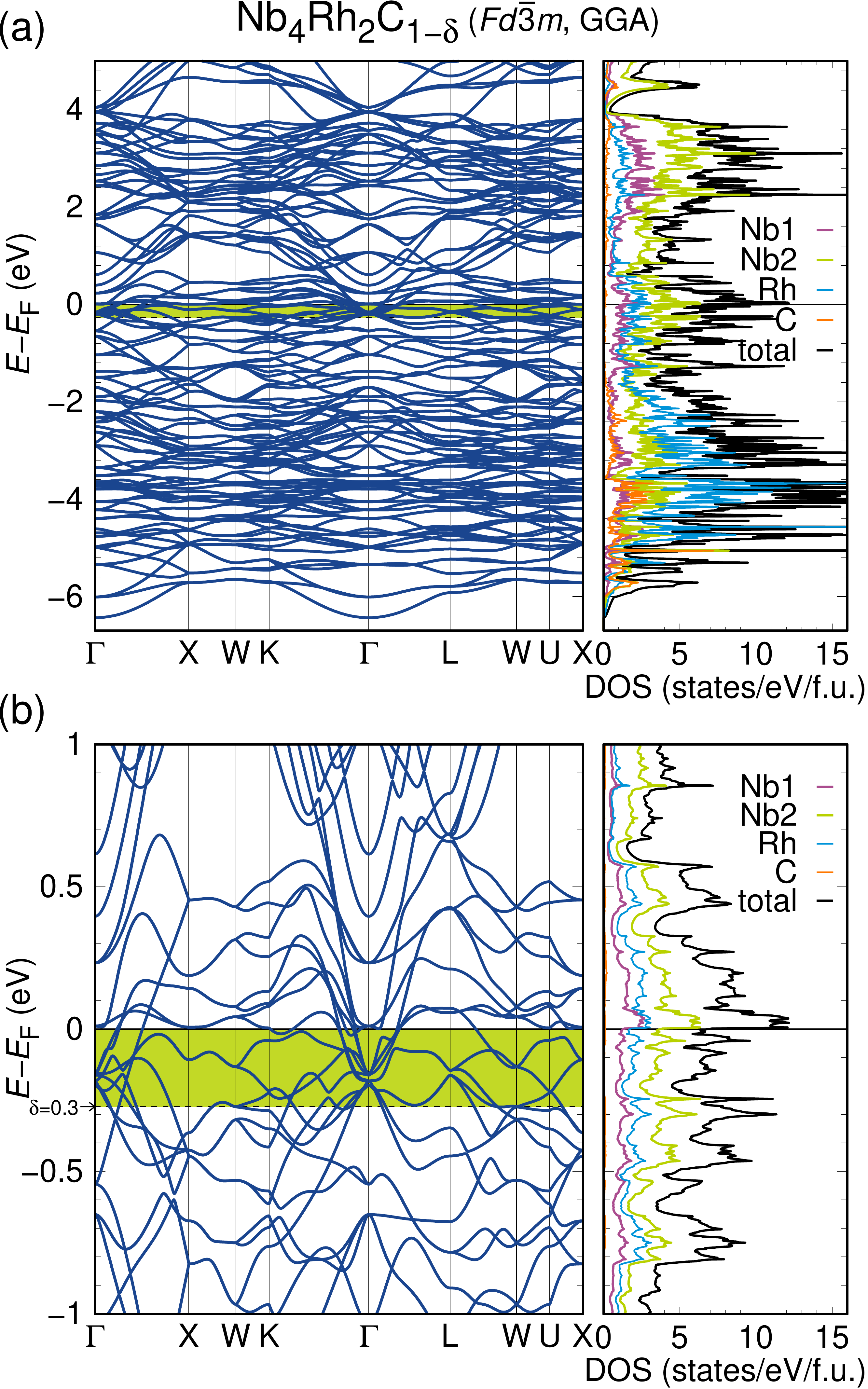}\\
\caption{Electronic structure of \ce{Nb4Rh2C_{1-\delta}} calculated with GGA. Overview (a) and detail (b) of the band structure and density of states. Green shading indicates the range $0\le \delta \le 0.3$.}
\label{fig:calc1}
\end{figure}

We have performed density functional theory calculations using the all electron full potential local orbital basis set~\cite{koepernik1999full} and the generalized gradient approximation (GGA) to the exchange correlation functional~\cite{perdew1996generalized}. These calculations were performed without and with spin-orbit coupling. We find that the effects of spin-orbit coupling in \ce{Nb4Rh2C_{1-\delta}} are moderate on the electronic band structure. The GGA calculation is shown in figure~\ref{fig:calc1}, for comparison the GGA+SO is provided in the Supporting Information. The green shading in figure~\ref{fig:calc1} indicates the range $0\le \delta \le 0.3$. The rather small influence of the spin-orbit coupling on the band structure is especially noteworthy, as a violation of the Pauli paramagnetic limit may be associated with strong spin-orbit coupling effect. There is, however, no large effect on the electronic structure observable as a result of spin-orbit coupling (SOC), despite the inclusion of the relatively heavy element Rh in this material.

The primitive cell of the $Fd\bar{3}m$ space group still contains four formula units of \ce{Nb4Rh2C_{1-\delta}} so that the bands in the large energy window shown in  figure \ref{fig:calc1}\,(a) correspond to $4d$ bands of 16 Nb and 8 Rh atoms. Carbon $2p$ density of states is below the Fermi level. There are no carbon states at the Fermi level. Hence, the dominant charge carriers are all metallic \textit{d}-bands from Nb and Rh. Integration of charges shows that approximately 2 carbon $2p$ states are unoccupied. This indicates a partial covalency of the C-Nb bonds. Furthermore, we find that Rh has approximately 8 occupied $4d$ states and Nb has approximately 3.5 occupied $4d$ states. 

In the Supporting Information, we show the Fermi surface for $\delta=0.3$ and $\delta=0$. The Fermi surface is found to be very three-dimensional. There are no obvious strong nesting vectors between the Fermi surface pockets that might hint towards a charge ordering in this system. Hence, no obvious electronic origin for the larger upper-critical field can be identified. 

\subsection{Conclusion}
In summary, we have shown that \ce{Nb4Rh2C_{1-\delta}} is a bulk superconductor with a critical temperature $T_{\rm c}$ = 9.75 K. \ce{Nb4Rh2C_{1-\delta}} was found to crystallize in the $\eta$-carbide structure. We find that \ce{Nb4Rh2C_{1-\delta}} is a bulk superconductor with a value for $\Delta C/\gamma T_{\rm c}$ of 1.64, close to the weak-coupling BCS value of 1.43. It is also an extreme type-II superconductor with a lower-critical field of $\mu_0 H_{\rm c1} $= 13.6 mT, and an upper-critical field $\mu_0 H_{\rm c2}(0)$ of 28.4 T, exceeding by far the weak-coupling BCS Pauli paramagnetic limit of $\mu_0 H_{\rm Pauli} =$ 18.1 T. In comparison to the two most commonly used commercial materials for superconducting magnets, NbTi and \ce{Nb3Sn}: This material has a critical temperature $T_{\rm c}$ close to NbTi, while displaying a much larger critical field $\mu_0 H_{\rm c2}(0)$ close to the one of \ce{Nb3Sn}. Expected causes for these superconducting properties may lay in an unusually large coupling of the electron spin and the SOC, or in an unconventional superconducting pairing process. The analysis of the crystal structure, the effect of SOC on the band structure, and the magnetic properties of the material, however, have offered no obvious explanation for the highly unusual upper-critical fields observed in this study. The determination of the origin of this effect will be of significant interest in future research.

\subsection{Supporting Information}
PXRD patterns of the varying sample compositions, ZFC-FC magnetization of the superconducting transition, normal state magnetization measurements, resistivity measurement between 2 K and 300 K, temperature-dependent specific heat capacities measurements between 2 K and 300 K, upper critical field fitting with the Ginzburg-Landau formalism, temperature-dependent magnetoresistance and Hall effect measurements between 2 K and 300 K, GGA+SO calculation, calculated Fermi surface with $\delta=0$ and $\delta=0.3$.

\begin{acknowledgement}

This work was supported by the Swiss National Science Foundation under Grant No. PZ00P2\_174015 and PCEFP2\_194183. The work at GUT was supported by the National Science Centre (Poland; Grant UMO-2018/30/M/ST5/00773). The authors thank Franck Krumeich for help with the TEM measurements. The Electron Microscopy was performed at the Scientific Center for Optical and Electron Microscopy (ScopeM). 

\end{acknowledgement}



\bibliography{etacarbide}

\end{document}